\documentclass[onecolumn,10pt]{article}
\usepackage[top=1.25in, bottom=1.25in, left=1.5in, right=1.5in]{geometry}
\usepackage{amsmath,amsfonts,amscd,amssymb}
\usepackage{graphicx}
\usepackage{overpic}
\usepackage{cancel}
\usepackage{rotating}
\usepackage{url}
\usepackage{caption}
\usepackage{color}
\usepackage{rotating}
\usepackage{multirow}
\usepackage{wrapfig}
\usepackage{mathtools}
\usepackage{subeqnarray}
\usepackage{setspace}
\usepackage{longtable,tabularx}
\usepackage{multicol}
\usepackage{palatino} 
\graphicspath{{figures/}}

\usepackage[bottom,flushmargin,hang,multiple]{footmisc}

\newcommand\blfootnote[1]{%
  \begingroup
  \renewcommand\thefootnote{}\footnote{#1}%
  \addtocounter{footnote}{-1}%
  \endgroup
}
\usepackage{hyperref}
\renewenvironment{abstract}
 {\par\noindent\rule{\linewidth}{.25pt}\par\noindent\textbf{\abstractname}\par\noindent \ignorespaces}
 {\par\noindent\medskip\rule{\linewidth}{.25pt}}
 
\newcommand{\be}{\mathbf{e}}

\newcommand{\bu}{\mathbf{u}}

\newcommand{\bx}{\mathbf{x}}

\newcommand{\bn}{\mathbf{n}}

\newcommand{\bU}{\mathbf{U}}

\newcommand{\bW}{\mathbf{W}}

\newcommand{\bbR}{\mathbb{R}}

\newcommand{\cN}{\mathcal{N}}
\newcommand{\cL}{\mathcal{L}}

\newcommand{\bxi}{\boldsymbol{\xi}}

\usepackage{titlesec}
\titleformat*{\section}{\bfseries}
\titleformat*{\subsection}{\itshape}
\titleformat*{\subsubsection}{\itshape}
\titleformat*{\paragraph}{\large\bfseries}
\titleformat*{\subparagraph}{\large\bfseries}

\title{\Large{\vspace{-.65in}{Prediction of Intermittent Fluctuations from Surface Pressure Measurements on a Turbulent Airfoil}}\vspace{-.25in}}
\author{\normalsize{Samuel H. Rudy$^*$ and Themistoklis Sapsis} \\ 
\footnotesize{Department of Mechanical Engineering, Massachusetts Institute of Technology, Cambridge, MA 02139} \vspace{-1 mm} \\
}
\date{}

\begin{document}
\maketitle

\blfootnote{$^*$ Corresponding author (shrudy@mit.edu).\\ \noindent \textbf{Software:}  \url{https://github.com/snagcliffs/Airfoil_EE}.}
\vspace{-.25in}
\begin{abstract}
This work studies the effectiveness of several machine learning techniques for predicting extreme events occurring in the flow around an airfoil at low Reynolds. For certain Reynolds numbers the aerodynamic forces exhibit intermittent fluctuations caused by changes in the behavior of vortices in the airfoil wake.  Such events are prototypical of the unsteady behavior observed in airfoils at low Reynolds and their prediction is extremely challenging due to their intermittency and the chaotic nature of the flow. We seek to forecast these fluctuations in advance of their occurrence by a specified length of time. We assume knowledge only of the pressure at a discrete set of points on the surface of the airfoil, as well as offline knowledge of the state of the flow. Methods include direct prediction from historical pressure measurements, flow reconstruction followed by forward integration using a full order solver, and data-driven dynamic models in various low dimensional quantities. Methods are compared using several criteria tailored for extreme event prediction. We show that methods using data-driven models of low order dynamic variables outperform those without dynamic models and that unlike previous works, low dimensional initializations do not accurately predict observables with extreme events such as drag.\\

\noindent\emph{Keywords--}
Extreme events,
unsteady aerodynamics,
machine learning
\vspace{-0.05in}
\end{abstract}


\section{Nomenclature}
\vspace{-8 mm}
{\renewcommand\arraystretch{1.0}
\noindent\begin{longtable*}{@{}l @{\quad=\quad} l@{}}
$\bu$, $p$  & Fluid velocity and pressure\\
$P$& Airfoil surface pressure time series\\
$C_d, C_l$& Drag and lift coefficients\\
$q$ & Smoothed drag coefficient \\
$\omega$ & Extreme event rate \\
$\tau$ & Load time for prediction of $q$ \\
$m$ & Mass matrix from spectral element grid \\
$w$ & Weights for proper orthogonal decomposition and flow reconstruction \\
$\mathcal{C}$ & chord-length \\
$\mathcal{L}_\bullet$ & Loss function used to train neural network $\bullet$ \\
$dt$ & Time step \\
$Re$ & Reynolds number \\
$u_\infty$ &  inlet velocity\\
$\Phi \Sigma \Psi^T$   & components of proper orthogonal decomposition of flow field \\
$\xi$ & Latent space representation of flow field in full-field neural network\\
$r$ & Rank of low dimensional representations of flow\\
$G$  & Network for estimating proper orthogonal decomposition time series \\
$E$  & Encoder network for flow reconstruction \\
$D$  & Decoder network for flow reconstruction \\
$F_{p/\psi / \xi}$  & Non-dynamic models for mapping $P / \psi / \xi$ to $q$ \\
$H_{p/\psi / \xi}$  & Dynamic models for variables $P$, $\psi$, and $\xi$\\
$S, R,F_1,\alpha$  & Precision, recall,  $F_1$ score, and area under S-R curve\\
$F_{1,opt}$  & Optimal test set $F_1$ score\\
$\alpha^*$  & Maximum adjusted area under precision-recall curve\\
\end{longtable*}}

\section{Introduction}

Extreme events are common features in engineering and scientific disciplines including climate, ocean engineering, and fluid structure interaction that are characterized by observables of a dynamical system exhibiting heavy tails \cite{sapsis2020statistics}. 
The outlier events populating these tails are of particular interest due to their effects on aerodynamics and fatigue, or other potentially adverse consequences.
However, the rarity and intermittency of such events also makes their prediction challenging.
There has been significant recent interest in sampling strategies \cite{mohamad2018sequential,sapsis2020output,blanchard2021bayesian}, optimization schemes \cite{blonigan2019extreme}, and tailored loss functions \cite{guth2019machine,qi2020using} for the prediction of extreme events. 
A common goal of many of the past works on extreme events and of the present work is the prediction of extreme events in advance of their occurrence.

The focus of this work is on the two-dimensional incompressible flow around an airfoil at low ($\mathcal{O}(10^4)$) Reynolds number. 
Dynamics of the flow around the airfoil at this Reynolds regime are highly nontrivial \cite{lissaman1983low} and have been shown to be characteristically different from those at higher Reynolds \cite{kim2011low}.
Previous works using both experimental and computational tools have found that slow moving airfoils exhibit a large range of wake behaviors, with qualitative changes in the nature of the flow occurring with small changes in angle of attack and Reynolds number \cite{menon2020aerodynamic, gopalakrishnan2017airfoil, wang2014turbulent}.
Similar unstable behavior has been observed in the flow around a cylinder, in the so called transitional regime between ordered and disordered behavior \cite{williamson1996vortex} as well as in vortex-induced-vibrations of flexible cylinders \cite{modarres2011chaotic}. 
Due to these instabilities, as well as an apparent lack of fidelity between computational and experimental results, it has been suggested that construction of rigid winged slow flying vehicles may be challenging if not impossible \cite{tank2017possibility}.  

Despite apparent challenges, there has long been considerable interest in the study of low Reynolds airfoils \cite{lissaman1983low}. 
In particular, recent works have explored the use of machine learning to estimate flow field and aerodynamic data from sensors on the surface of the airfoil.
These include methods for flow reconstruction from limited sensors using neural networks \cite{maulik2020probabilistic, gomez2019unsteady}, filtering based flow estimation \cite{de2017enkf,da2020flow},  as well as prediction of aerodynamic coefficients \cite{dawson2015data}. 
Deep learning has also been used for estimating properties of the flow used in low order vortex models. 
In \cite{hou2019machine,le2020deep}, authors use neural network based methods to estimate the leading edge suction parameter(LESP).  
The model studied in this work has a Reynolds number of 17,500, substantially higher than other computational works which focus on Reynolds number in the range of $\mathcal{O}(10^2) - \mathcal{O}(10^3)$ \cite{le2020deep, de2017enkf, da2020flow, maulik2020probabilistic}.
This is closer to the lower end of the range considered by experimental work \cite{dawson2015data, gomez2019unsteady}. 
Unlike some other works \cite{dawson2015data, gomez2019unsteady, hou2019machine, le2020deep}, this work does not study the effects of pitching motions or disturbances in the incident velocity. 
We instead focus on prediction of intermittent fluctuations in the aerodynamic coefficients of a stationary airfoil, which have not been the focus of previous works.

This work applies several machine learning techniques to predict fluctuations in the drag coefficient of an airfoil in the transitional regime where we observe chaotic and intermittent behavior.  It follows a broader trend of the application of tools from machine learning to problems in fluid dynamics.  For a broader view, the interested reader may refer to a number of recent articles outlining and discussing the role of machine learning in fluid dynamics; \cite{brunton2020machine} provides an excellent overview of many of the applications machine learning has seen in fluids, \cite{fukami2020assessment} provides an assessment of several common supervised learning methods applied to flow reconstruction, super-resolution, and coefficient estimation, and \cite{brenner2019perspective} provides a discussion of the role of machine learning in fluids, as well as some pitfalls and concerns. 

The paper is organized as follows: In section \ref{sec:problem_description} we formulate the problem of predicting aerodynamic fluctuations. 
Section \ref{sec:methods} describes the methodology used in this work, including flow reconstruction methods in section \ref{subsec:flow_reconstruction}, forecasting methods for aerodynamic fluctuations in section \ref{subsec:aero_forecast} and a discussion of performance metrics in section \ref{subsec:error_metrics}.  Results are shown in section \ref{sec:results} with comparisons between all methods.  In section \ref{sec:discussion} we offer closing thoughts, pitfalls, and potential future directions based on this paper. 

\section{Problem description} \label{sec:problem_description}

We consider a NACA 4412 airfoil at chord-length based Reynolds number of $Re=17500$ and $5^\circ$ angle of attack.  Flow around the airfoil is simulated using the spectral element code Nek5000 \cite{nek5000-web-page} according to the incompressible Navier-Stokes equations given by,
\begin{equation}
\begin{aligned}
\dfrac{\partial \bu}{\partial t} + \bu \cdot \nabla \bu &= - \nabla p + \dfrac{1}{Re}\nabla^2 \bu \\
\nabla \cdot \bu &= 0 .
\end{aligned}
\label{eq:incomp_ns}
\end{equation}

The computational grid uses 4368 elements with spectral order 7.  Statistics of observables relevant to this work were found to be in agreement with those from a shorter simulation using a more resolved grid having 14144 elements.   A convective boundary condition is used for the outflow \cite{dong2015convective}. 
The spectral element grid without Gauss-Lobatto Legendre interpolation points and a snapshot of the vorticity are shown in the top row of Figure \ref{fig:grid_vort}.  
Further details of the numerical simulation are provided in Appendix A and software for reproducing data used in this work is available online.

\begin{figure*}
\centering
\includegraphics[width=\textwidth]{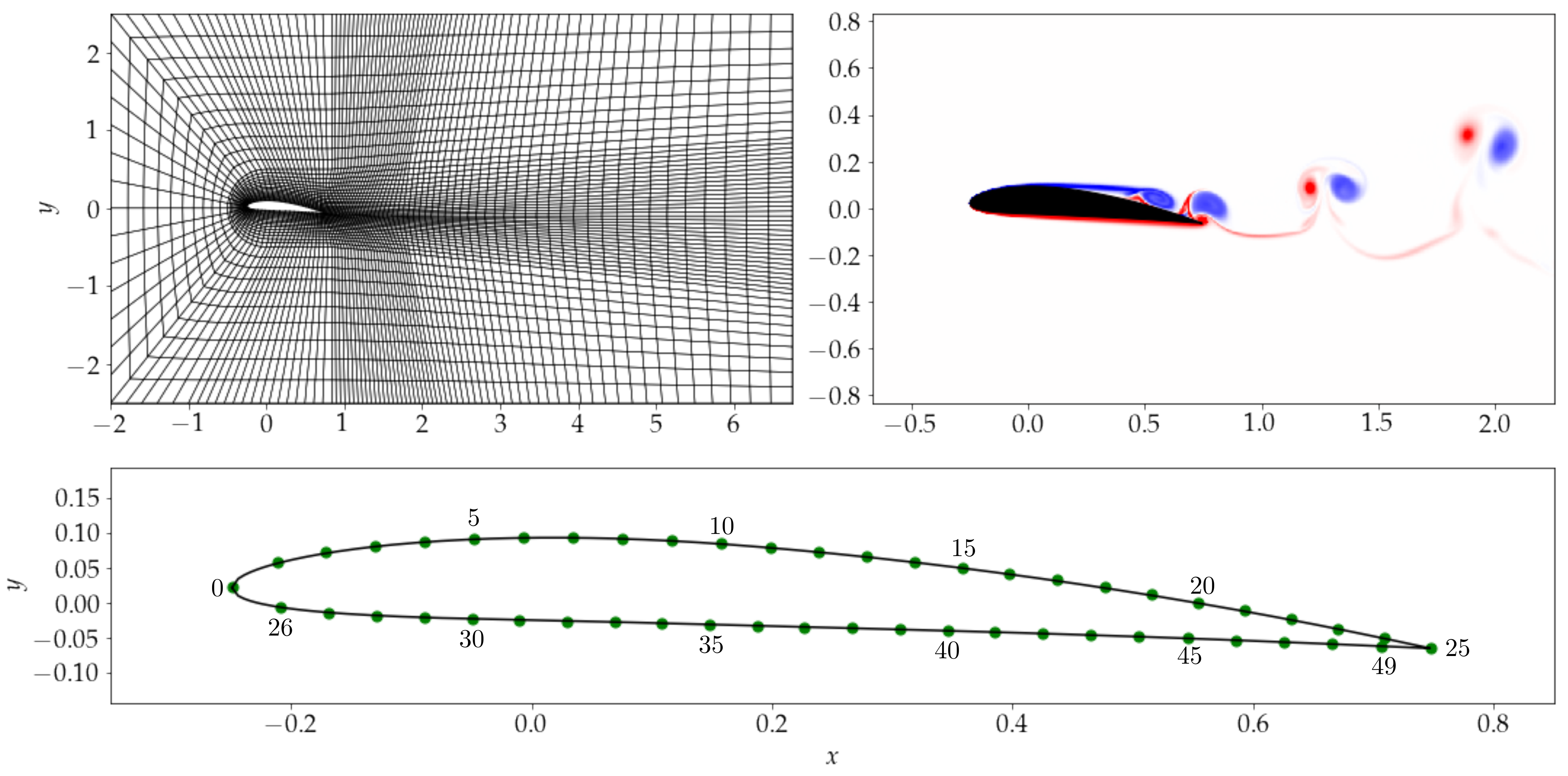}
\caption{Left: Domain of computational problems showing outlines of spectral elements without internal interpolation points.  Right: Snapshot of vorticity. Bottom: Sensor placement showing indexing from 0-25 along top and 25-49 along bottom of airfoil.  Axes on bottom figure not drawn to scale.}
\label{fig:grid_vort}
\end{figure*}
Pressure recordings along the surface of the airfoil $P(t)$ are taken at a discrete set of 50 points around the airfoil at intervals of $dt=0.01$ throughout the simulation.
These locations are shown in the bottom panel of Fig. \ref{fig:grid_vort}.
Aerodynamic force is computed using the pressure and skin friction.  
This decomposes into the streamwise $(x)$ and  cross-flow $(y)$ directions, defined by:
\begin{equation}
\vec{F}(t) = \oint \boldsymbol{\tau}(t) - p(t)\bn \, ds = D(t) \be_x + L(t) \be_y
\label{eq:aerodynamic_force}
\end{equation}
where $\boldsymbol{\tau}$, $p$, and $\bn$ are the skin shear stress, pressure, and wall normal vector, and the integral is taken over the airfoil surface.  Forces are then used to compute the non-dimensional drag-coefficient $C_d$ and lift coefficient $C_l$, defined as:
\begin{equation}
C_d(t) = \dfrac{2D(t)}{\rho u_\infty^2 C},\hspace{1 cm} C_l(t) = \dfrac{2L(t)}{\rho u_\infty^2 C},
\label{eq:aerodynamic_coefficients}
\end{equation}
where choord length $C=1$, density $\rho=1$, and free-stream velocity $u_\infty = 1$.  

\begin{figure*}
\centering
\includegraphics[width=\textwidth]{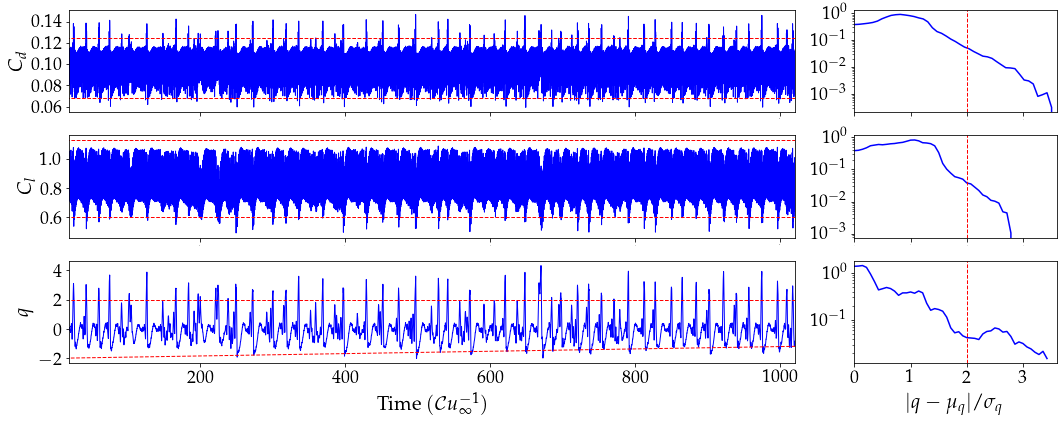}
\caption{Left: Aerodynamic coefficients $C_d$ and $C_l$, and smoothed drag coefficient $q$ as defined in Eq. \eqref{eq:q_definition}.  Red dashed lines indicate $\pm 2$ standard deviations away from the mean.  Right: Histograms of the deviation of each quantity showing typical heavy tails of systems with extreme events.}
\label{fig:QOI_fig}
\end{figure*}

The two dimensional simulation yields a quasi-stable behavior in which intermittent fluctuations are observed in the aerodynamic coefficients, shown in the first two rows of Figure \ref{fig:QOI_fig} alongside the density functions of their absolute deviations.
We note that the density functions clearly exhibit the expected ``heavy tails'' associated with observables of dynamical systems with extreme events \cite{sapsis2021statistics}.
This regime of intermittent fluctuations is persistent for the entirety of the simulation used in this work, but with altered conditions may exhibit mode switching to a state with more regular oscillations. 
Further details of this case are given in Appendix C.
In this work we focus solely on the regime where intermittent fluctuations are observed.

The goal of the present work is to predict these intermittent fluctuations in the aerodynamic coefficients using information regarding the surface pressure of the airfoil in advance of their occurrence by some lead time $\tau$.  
To focus on non-periodic behavior, we consider predictions on a smoothed time series derived from the drag coefficient.
Specifically, 
\begin{equation}
q(t) = \left( C_d (s) \ast \cN\left(s \left| 0,\frac{1}{\left(2f_{peak}\right)^2} \right.\right) \right) (t),
\label{eq:q_definition}
\end{equation}
where $f_{peak} \approx 1.44$ is the peak frequency of the drag coefficient.  
In practice, the convolution in Eq. \eqref{eq:q_definition} is computed using a compactly supported kernel having width $3/f_{peak}$.
The time series $q(t)$ is also normalized to be mean zero and unit variance.
The quantity $q(t)$ captures the non-periodic behavior of the drag coefficient, in particular maintaining extreme events and the heavy tailed deviation.  
The goal of this work is thus concisely stated as learning data-driven models for the prediction problem,
\begin{equation}
P(s \leq t) \to q(t+\tau),
\label{eq:prediction_problem}
\end{equation}
at various $\tau \geq 0$. 

\section{Methods} \label{sec:methods}

In this section we formulate several data-driven models for Eq. \eqref{eq:prediction_problem}.  
We assume knowledge of measurements of pressure at a discrete set of points along the surface of the airfoil up to the current time $t$ and seek to predict the value of the extreme event indicator at time $t+\tau$.
We denote the time series for pressure measurements up to the current time $t$ as $P_t = P(s \leq t)$.
The minimal problem is thus learning a function directly from historical pressure measurements to the $q(t+\tau)$, which may be done using a variety of standard machine learning tools.

\begin{figure*}[h]
\centering
\includegraphics[width=\textwidth]{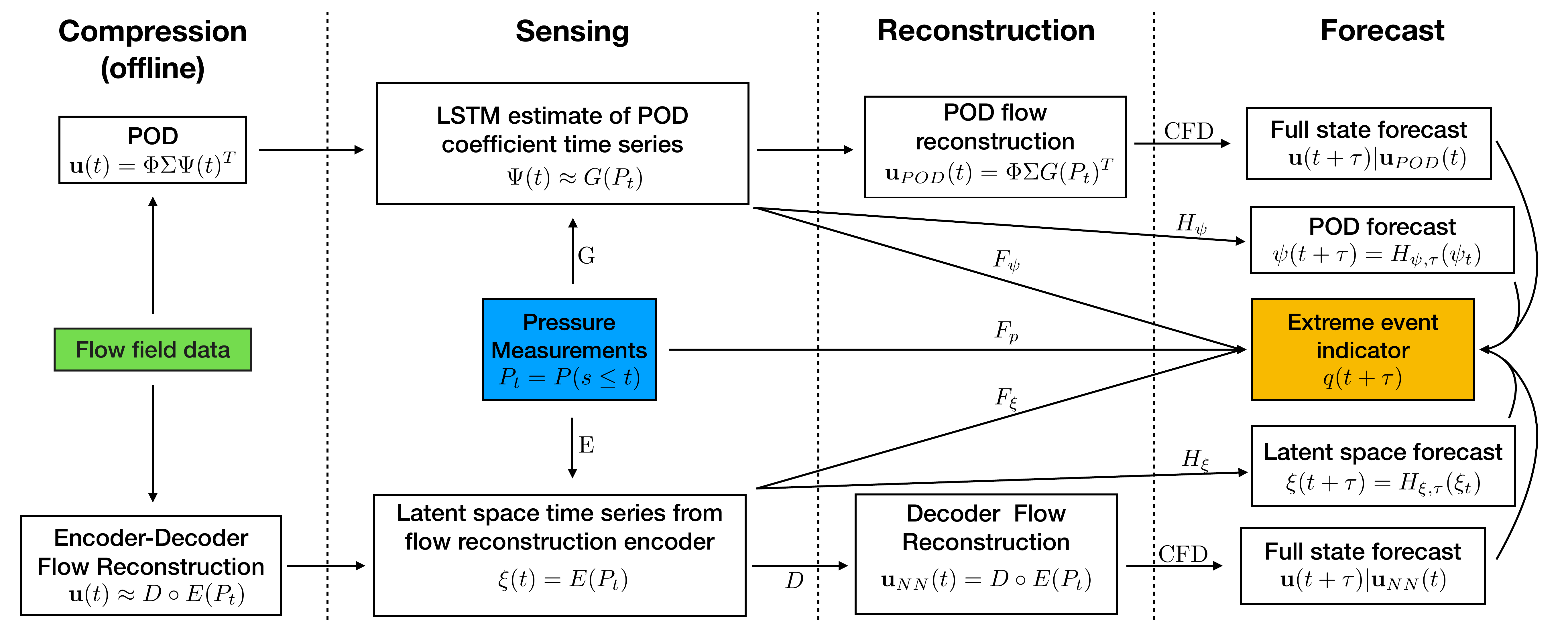}
\caption{Schematic of methods for EE prediction showing various flows of information from assumed knowledge of historical pressure measurements $P_t$ to future value of quantity of interest $q(t+\tau)$.  Dynamic model for pressure measurements has been omitted.}
\label{fig:airfoil_schematic}
\end{figure*}

A baseline predictor for Eq. \eqref{eq:prediction_problem} is found using a standard recurrent neural network, such as a long-short-term-memory networks (LSTM) \cite{hochreiter1997long}, to interpolate a function directly from $P_t$ to $q(t+\tau)$.
Alternatively, we may try to improve forecasts of $q(t)$ through the offline use of flow field data.
Previous works \cite{blonigan2019extreme} have used modal representations of a flow field  combined with adjoint equations  to learn precursor states to extreme events. 
In this work we discuss two methods for compressing flow field data, sensing expansion coefficients in the compressed basis, and exploiting this knowledge for potentially improved prediction of $q(t+\tau)$: the proper orthogonal decomposition and neural network based flow reconstruction.
In each case we study the predictive capability of initializing a flow solver with the reduced order initial condition, predicting directly from historical representations in the reduced space, and neural network based reduced order models.
We also consider data-driven dynamic models for the pressure measurements.

The methods considered in this work are summarized in Figure \ref{fig:airfoil_schematic}.  
Methods are separated into an offline compression stage, a sensing stage where we infer the reduced order state from point pressure measurements, reconstruction of the flow field from the reduced order state, and finally forecast of the quantity of interest, $q(t)$.
Learned functions $G$, $E$, and $D$ are predictors of the POD mode time series, latent space encoder, and flow reconstruction network, respectively.  
The letter $F$ has been used to denote LSTM predictions that do not use a dynamic model, with subscript indicating input.
$H$ is used to denote neural network based reduced order models.
In each case, we use time series for pressure at discrete points on the surface of the airfoil as the starting point of the online prediction.  

Details on each method are provided in the following sections.
We assume familiarity with common implementations of neural networks and stochastic optimization.  
In particular, the work in this manuscript makes frequent use of deep LSTMs \cite{hochreiter1997long} and the Adam method for optimization \cite{kingma2014adam}.
The unfamiliar reader may find an excellent reference in the free online textbook \cite{goodfellow2016deep}. 
Further details on the structure and implementation of neural networks is given in Appendix B.

\subsection{Flow field compression and reconstruction}\label{subsec:flow_reconstruction}

We begin with a discussion of the two offline methods for flow reconstruction: the proper orthogonal decomposition (POD) and an LSTM based encoder-decoder pair.  
In this sub-section we provide an overview of the methodology used to form the reduced rank representation for each of these two cases as well as methods for approximating time series associated with each POD mode.

\subsubsection{Proper orthogonal decomposition and sensing}\label{subsub:PODFR}

The proper orthogonal decomposition is a standard tool for decomposing a flow field into spatial modes that are orthogonal with respect to a given inner-product and whose time evolution are also orthogonal \cite{holmes2012turbulence,taira2017modal}.  We apply the POD to the velocity field around the airfoil using a weighted inner product.  Specifically, the POD finds matrices $\Phi$, $\Sigma$, and $\Psi$ which are discretized solutions to,
\begin{equation}
\bu (\bx,t) - \overline{\bu}(\bx) = \Phi(\bx) \Sigma \Psi(t)^T,
\label{eq:POD}
\end{equation}
where,
\begin{equation}
\begin{aligned}
& \Sigma = diag(\sigma_1, \hdots, \sigma_r), \,\, \sigma_i \geq \sigma_{i+1}\geq 0\\
&\langle \phi_i, \phi_j \rangle_w = \int_\Omega \phi_i(\bx)^T\phi_j(\bx)w(\bx)\,dx = \delta_{i,j} \\
&\langle \psi_i, \psi_j \rangle_w = \int_0^T \psi_i(t)\psi_j(t)\,dt = \delta_{i,j},
\end{aligned}
\label{eq:POD_constraints}
\end{equation}
and $w(\bx) > 0$ is a weight used to focus the inner product, and thus variation explained by POD, on regions of interest near the surface of the airfoil.  A similar weighted approach was used in \cite{blonigan2019extreme} where a wall focused POD was used as a basis for predicting extreme dissipation events in channel flow.  We use the family of weights given by,
\begin{equation}
w(\bx) = \frac{1-\epsilon}{1+e^{(d(\bx) - l)/\delta}} + \epsilon
\label{eq:weights}
\end{equation}
where $d(\bx)$ is the distance from $\bx$ to the surface of the airfoil.  Equation \eqref{eq:weights} describes a smooth sigmoidal curve that decays from 1 (assuming $l \gg \delta$) at the airfoil surface to $\epsilon$ in the far field.  For small $\delta$, this transition is localized around $d=l$ and the weights are approximately 1 for $d < l - \delta$ and $\epsilon$ for $d > l + \delta$.  We use parameter values $l=1$, $\delta=0.1$, and $\epsilon = 0.1$.  The POD is therefore principally focused on describing variation in the velocity field within one chord length of the airfoil surface, with approximately one tenth the weighting for variation outside this region.

In the case where $w(\bx)=1$ the POD is equivalent to the singular value decomposition of the mean subtracted data, also knwon as principal component analysis. 
For non-identity weights, $\Sigma$ and $\Psi(t_j)$, $j=1,\hdots, m$ are given by the eigenvalue decomposition of $\bU^T\bW_m\bU$ where $\bar{\bU}\in \bbR^{n\times m}$ is the mean-subtracted velocity data and $\bW_m$ is a diagonal matrix with $w(\bx_i) m(\bx_i)$ along the diagonal where $m(\bx_i)$ is the mass associated with that grid point for the spectral element grid \cite{holmes2012turbulence}. 
Modes $\Phi$ are subsequently computed using their definition in Eq. \eqref{eq:POD}. 
Alternative methods may compute $\Phi$ before $\Sigma$ and $\Psi$, but these suffer from numerical issues for $\epsilon \ll 1$.

\begin{figure*}[h]
\centering
\includegraphics[width=\textwidth]{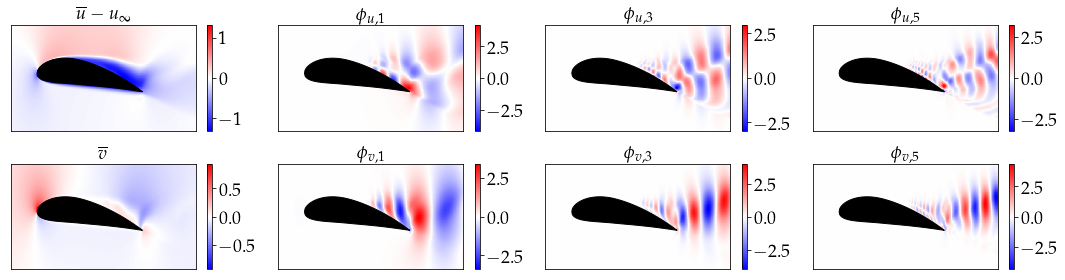}
\caption{The mean velocity field and three POD modes.  Axes not drawn to scale.}
\label{fig:POD_modes}
\end{figure*}

\begin{figure*}[h]
\centering
\includegraphics[width=\textwidth]{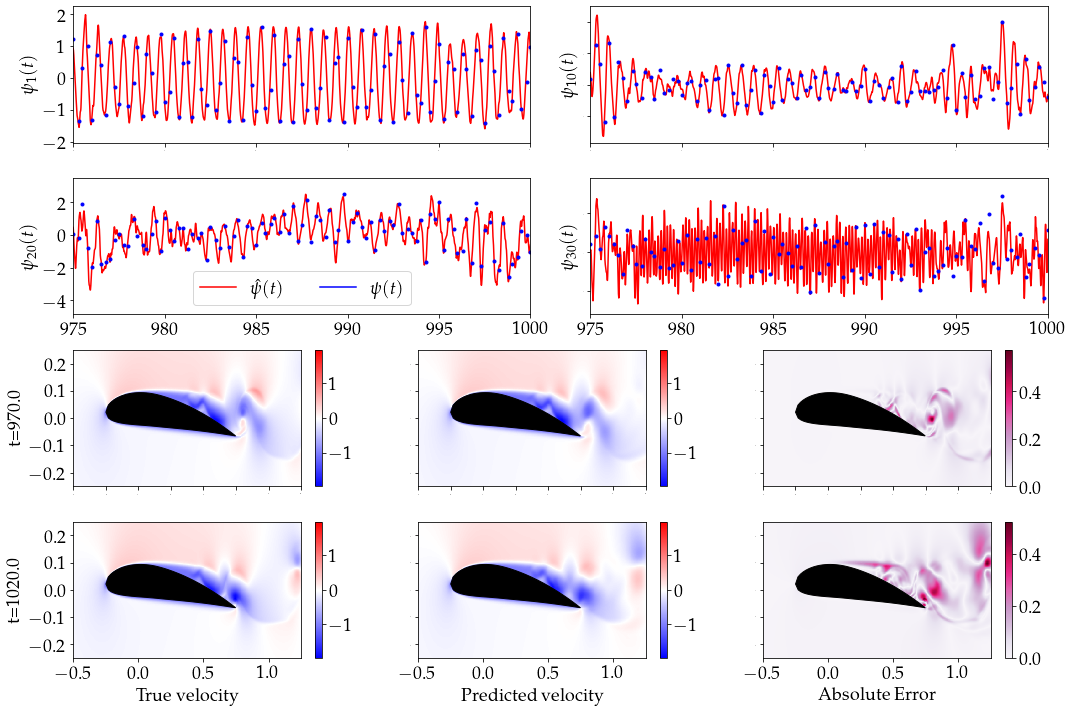}
\caption{Flow field reconstruction using LSTM to predict time series for first 32 POD modes using pressure measurements, $P_t$.  Top [Operator $G(P_t)$]: LSTM prediction of POD time series for 1$^{st}$, 10$^{th}$, 20$^{th}$, and 30$^{th}$ modes (red) and sparser true values (blue).  Bottom [Operator $\Phi \Sigma G$]: POD reconstruction of streamwise velocity compared to two true fields within test set data.}
\label{fig:POD_reconstruction}
\end{figure*}

Applying POD to the airfoil data yields modes $\phi_i$, shown in Fig. \ref{fig:POD_modes}, singular values $\Sigma$, and time series $\psi_i(t)$ corresponding to each mode $\phi_i$.  In the online phase of any prediction method, we will only have access to $P_t$, not $\psi_i(t)$.  The latter may be estimated from sparse or gappy measurements \cite{venturi2004gappy}.  We therefore train a deep LSTM model to estimate the current POD representation of the flow from pressure measurements.  Letting $\psi_r(t)$ be the rank $r$ truncation of the time series of the POD expansion we have,
\begin{equation}
\hat{\psi}_r(t) = G(P_t).
\label{eq:G_P_to_Psi}
\end{equation}
The exact form of and training procedure for $G$ is described in greater detail in Appendix B.
True values of POD coefficients as well as there estimates from pressure via Eq. \eqref{eq:G_P_to_Psi} are shown in Fig. \ref{fig:POD_reconstruction}.  Note that reconstructed time series are filled in on a much denser grid than true values since they are computed from the finely sampled pressure time series.  
Temporal resolution on $\psi(t)$ is limited by the number of output files saved during numerical simulation as well as memory limitations in the computation of the POD.

It is worth noting that the time series $\psi(t)$ are normalized to unit variance and $\Sigma$ is not considered in loss function. 
Hence, error in higher modes is treated the same as error in lower modes. 
Higher modes did still have higher error, perhaps because they tended to exhibit more chaotic behavior. 
The authors did not explore the loss function exhaustively since doing so would be a significant research endeavor on its own.

After prediction of $\hat{\psi}(t)$ using $G$ one may reconstruct an approximation of the full flow field using Eq. \eqref{eq:POD}. 
The lower two rows on Fig. \ref{fig:POD_reconstruction} show the results of this in the streamwise direction as well as the true velocity and absolute error for two snapshots taken from the testing dataset, i.e. snapshots not seen by the optimization algorithm used to learn $G$.  These reconstructions suffer from multiple sources of error.  Expanding the difference between true and reconstructed fields, we get,
\begin{equation}
\begin{aligned}
\bu(t) - \hat{\bu}_{POD}(t) &= \Phi\Sigma\psi (t)^T - \Phi_r\Sigma_r G(P_t)^T \\
&= \underbrace{\Phi_{-r}\Sigma_{-r}\psi_{-r} (t)^T}_{\text{Unresolved}} - \Phi_r\Sigma_r (\underbrace{\psi_r(t) - G(P_t)}_{\text{NN Error}} )^T,
\end{aligned}
\label{eq:POD_reconstruction_error}
\end{equation}
where the subscript $-r$ denotes those components not in the first $r$.  Error in the neural network increases for modes with high frequency and chaotic behavior, such as mode 30 shown in Fig. \ref{fig:POD_reconstruction}.
However, significant error is also incurred from unresolved modes due to the slow decay of singular values $\Sigma$. 
While spatially periodic translational behavior may be represented with pairs of modes (see Eq. (34) in \cite{taira2017modal}) the system studied in the present work tends to shed isolated pairs of vortices. 
Moreover, it also exhibits intermittency, which may be difficult to capture in a POD basis.
The authors are not aware of efficient linear methods for representing translation of sparse structures such as the wake vorticies observed in this data.

\subsubsection{Neural network flow reconstruction}\label{subsub:NNFR}

In light of some of the deficiencies of the POD and motivated by the successful applications of neural networks to problems in fluid dynamics  \cite{milano2002neural,lu2021learning,raissi2019physics,brunton2020machine,fukami2020assessment} we also consider neural network based approaches to flow reconstruction.  
The approach in this work is to use an LSTM-based encoder coupled with a fully connected network predicting the velocity at each grid point.  
The fully connected decoder resembles that used in \cite{erichson2020shallow} and \cite{maulik2020probabilistic} though do to the large computational grid we have not used the probabilistic formulation considered in the later.  
We note however, that neither of the aforementioned papers included the use of history terms in their prediction of the fluid state, as we will show is done by the LSTM encoder for the networks considered in this work.
For brevity, we will call networks of this type full field neural networks (FFNN), indicating that the output of the network is the values of fluid velocity at each grid point used by a solver. 

As in the POD case, we represent the fluid velocity around an airfoil at time $t$ using a low dimensional representation, $\bxi(t) \in \bbR^r$.  The time history of the pressure sensors is encoded to the state using an LSTM given by,
\begin{equation}
\bxi(t) = E (P_t).
\label{eq:LSTM_encoder}
\end{equation}
Since the full state of the flow field is encoded in $\bxi(t)$, one may tune the dimension of $\bxi$ to acquire a desired rank for the reduced order representation of $\bu$.  
We found improvements in reconstruction accuracy up to approximately rank $r=32$, with minimal improvement at high values.  
We therefore use $r=32$ for the remainder of this work unless noted otherwise.
The reconstructed velocity field is then given by,
\begin{equation}
\hat{\bu}(\bx_i, t_m) = D(\bxi)_i,
\label{eq:grid_decoder}
\end{equation}
where $D$ is a standard fully connected neural network with the final layer being linear.  Taken together, Eq. \eqref{eq:LSTM_encoder} and \eqref{eq:grid_decoder} form a recurrent network from pressure sensor time histories to the fluid velocity given by,
\begin{equation}
\hat{\bu}(x_i, t_j) = D \circ E (P(t_m))_i,
\label{eq:nn_composition}
\end{equation}
which are trained together using numerical simulation data.  Further details on the network structure may be found in Appendix B.

We train the networks using a weighted square-error loss function designed to favor accurate reconstruction near the wall boundary;
\begin{equation}
\begin{aligned}
\cL_{E,D} &= \sum_j \int_\Omega \left\| u(\bx, t_j) - \hat{u}(\bx, \bxi) \right\|^2 w(\bx) \, dx,
\end{aligned}
\label{eq:cotinuous_net_loss}
\end{equation}
where $\Omega$ is the computational domain and $w(\bx)$ is as defined in Sec. \ref{subsub:PODFR} using $\epsilon = \delta = 0.1$ and $l = 1$.  It is also possible to weigh the loss function to more heavily penalize errors in times immediately preceding an extreme event, but doing so was observed to have little effect on prediction performance.  In practice, the integral in Eq. \eqref{eq:cotinuous_net_loss} is approximated using the mass matrix $m$ obtained from the spectral element grid.  The expression simplifies to a simple weighted sum of squares error given by,
\begin{equation}
\begin{aligned}
\cL_{E,D} &\approx \sum_j \sum_i \left\| \bu(\bx_i, t_j) - \hat{\bu}(\bx_i, \bxi) \right\|^2 w(\bx_i) m(\bx_i).
\end{aligned}
\label{eq:discrete_net_loss}
\end{equation}

\begin{figure*}[h]
\centering
\includegraphics[width=\textwidth]{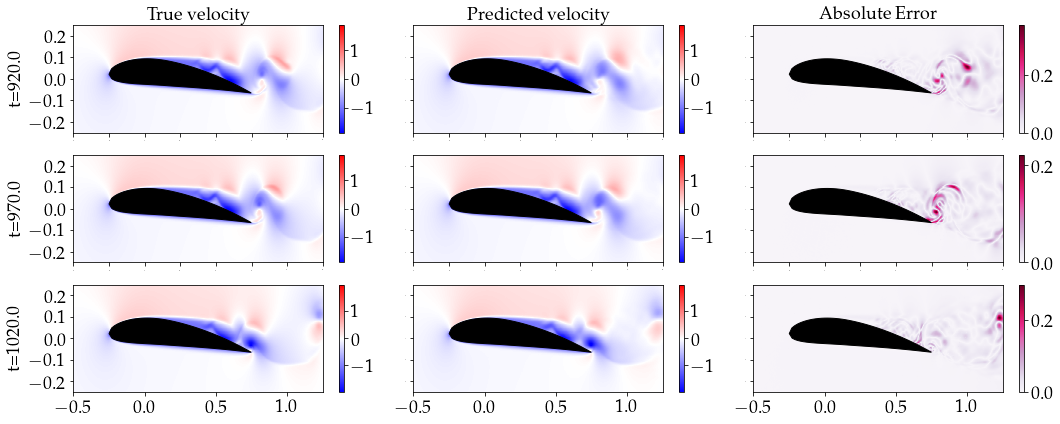}
\caption{ [Operator $D \circ E (P_t)$] Reconstruction of streamwise velocity ($u_\infty = 1$ subtracted) using FFNN for flow compression with 32 dimensional latent space and LSTM to map the pressure measurements, $P_t$, to the latent variables, $\bxi$.  Note maximum error is approximately half that of POD, and regions of high error are significantly more localized.}
\label{fig:FFNN_reconstruction}
\end{figure*}

Snapshots of the true streamwise velocity, the FFNN predicted streamwise velocity, and absolute error are shown in Fig. \ref{fig:FFNN_reconstruction}, which includes the same fields shown in Fig. \ref{fig:POD_reconstruction} to illustrate POD reconstruction.  
Note that the maximum absolute error using neural networks is considerably lower than that of POD and that regions of high error are more localized.
This clearly indicates the superior performance of FFNN to POD for flow reconstruction in this particular case.

The choice to use a very large output layer of the neural network makes predictions specific to the particular grid used in training, though interpolation schemes could be used in other cases.
The FFNN decoder's size also makes it highly memory intensive, which limits batch size in training.  
It is plausible that similar networks for three dimensional flows would require coordinate descent like algorithms where fractions of the output weights are updated on any given batch.
This is in contrast to operator type networks, where spatial coordinate $x$ is given as input \cite{lu2021learning}.
Methods based on the latter were implemented without physical constraints for flow reconstruction from pressure measurements, but were found to underperform the full-field neural networks discussed in this work.  
We note that this could be in part due to using a neural encoder of pressure measurements rather than sparse function evaluations, as used in \cite{lu2021learning}.
It is also possible that the use of physics-informed methods could improve prediction accuracy of the operator type networks and this is noted as a potentially interested research question.  
However, we consider such an approach to be outside of the scope of this work.

\subsection{Forecasting aerodynamic fluctuations}\label{subsec:aero_forecast}

We now consider online methods for Eq. \eqref{eq:prediction_problem}.  
We separate these methods into two broad categories; those that do not use dynamic models, and those that use dynamic models such as the Navier Stokes equations or data-driven dynamic models. 

\subsubsection{Non-dynamic methods}\label{subsec:single_step}

We first consider methods that do not employ any sort of dynamic model.  These are simply interpolations a function from the time history of an input quantity to the future value of $q(t)$.  
The general form is given by,
\begin{equation}
q(t + \tau) = F_\bullet (\bullet_t),
\label{eq:general_single_step}
\end{equation}
where $\bullet \in [P, \psi, \xi]$ and $F$ is a deep LSTM mapping some time series of historical measurements to the future quantity of interest.  In the case where the reduced representation of the flow state is used we have:
\begin{equation}
\begin{aligned}
q(t + \tau) &= F_\psi \circ G (P_t) \\
q(t + \tau) &= F_\xi \circ E (P_t),
\end{aligned}
\label{eq:non_dynamic_feature_maps}
\end{equation}
so the POD sensing network $G$ and neural network encoder $E$ may be considered as feature maps for the forecasting networks $F_{\psi / \xi}$.  Each of $F_{P/\psi/\xi}$ are trained using the mean square error loss function.  Other loss funcitons more specific to extreme events were considered but found to make negligible difference to the resulting trained network.

Non-dynamic methods of the form given by Eq. \eqref{eq:general_single_step} may be favorable for several reasons. 
They are simple, easy to train, and based on the ubiquitous and highly effective LSTM network structure.
They may be of particular interest due to the computational savings offered by avoiding dynamic models. 
Approximation of $q(t+\tau)$ may be rapidly computed from the state of the LSTM, which is updated online from streaming data.
However there are also downsides to the non-dynamic approach.
As $\tau$ becomes large, we are approximating larger steps of a chaotic time series. 
The problem thus becomes very sensitive to inputs, and balancing sufficient model complexity with overfitting becomes challenging.
Moreover, prediction of $q(t+\tau)$ for any given lead time $\tau$ requires its own trained network.  

\subsubsection{Full order dynamical model based prediction methods}\label{subsub:CFD_prediction}

Previous work predicting extreme events for turbulent flows has used a low dimensional representation of the flow as initial condition for a fluid simulation and employed the adjoint to evaluate the gradient of the future quantity of interest with respect to coefficients in the low dimensional expansion \cite{blonigan2019extreme}. 
This is possible in cases where the low order initial condition is sufficiently close to the full order state to track its behavior.
Here we evaluate if this is the case for low dimensional initialization from both the POD basis and FFNN.
Numerical simulations are initialized using either the FFNN based reconstructed velocity or the low dimensional reconstruction using the known POD modes and singular values, along with the estimated temporal coefficients $\hat{\psi}(t) = G(P_t)$.
Future values of $q(t)$ are computed using the same convolution as in Eq. \eqref{eq:q_definition};
\begin{equation}
\begin{aligned}
\hat{q}(t+\tau) &= \left( \hat{C}_d(s; \hat{\bu}) \ast \cN\left(s \left| 0,\frac{1}{\left(2f_{peak}\right)^2} \right.\right) \right) (t+\tau) \\
\text{where, } &\hat{\bu} (t) = D(\xi) \text{  or  } \hat{\bu} (t) = \Phi \Sigma G(\xi),
\end{aligned}
\label{eq:CFD_forecast}
\end{equation}
where $\hat{\bu}(s), s\in[t,t+\tau] $ is found via numerical integration of the Navier-Stokes equations.
Since the time series for $\hat{C}_d$ is much shorter than the full simulation used for training neural networks, the common practice of zero padding distorts the values of $\hat{q}$ near the endpoints. 
We instead truncate and re-normalize the smoothing kernel. 
For valid comparison, the same is done for the true drag coefficient truncated to the interval $[t, t+\tau]$. 
Following Eq. \eqref{eq:CFD_forecast}, time series for $q$ and $\hat{q}$ are normalized using the mean and variance of the full length time series defined in Eq. \eqref{eq:q_definition}.

\subsubsection{Data-driven dynamical model based prediction methods}\label{subsec:xi_rom}

We also consider forecasts of either i) the pressure signal, $P(t)$, or ii) the reduced order state of the fluid flow, using data-driven dynamic models. 
In the later case we formulate and evaluate reduced order dynamical models for both the reduced POD state and the latent representation learned by the FFNN. 
For the POD ROM, we train the dynamic model on the estimated POD coefficients ($\hat{\psi} = G(P_t)$), instead of the true coefficients $\psi$, since the former may be found exactly from airfoil surface pressure.
This eliminates error due to imprecise initial conditions.
Learning a data-driven model of the pressure signal along the surface of the airfoil has, to the best of the authors knowledge, not been used in previous works.

Forecasting models in each case are LSTM networks mapping historical measurements of a given quantity ($P/\hat{\psi}/\xi$) to its value $\kappa$ timesteps in the future.  They are represented as,
\begin{equation}
P(t_{j+\kappa}) = H_P (P(t_j), P(t_{j-\kappa}), \hdots)),
\label{eq:basic_lstm}
\end{equation}
with similar networks for $\hat{\psi}$ and $\xi$.
Here $\kappa$ is taken to be three in order to alleviate some of the numerical difficulties with training data-driven dynamic models in the small timestep limit \cite{levine2021framework}.
Thus, the LSTM maps historical measurements of the dynamic quantity $P$ to the value of that quantity three steps in the future: $P(t_{j+3})$.
Since $P$ is measured every $dt=0.01$ time units, the LSTM is effectively a dynamic model for $P$ with timestep equal to $\kappa \, dt = 0.03$.
Networks are trained using the mean square error loss over a prediction window of 20 steps;
\begin{equation}
\mathcal{L}_{H_P} = \sum_i \sum_{j=1}^{20} \| H_P^j(P_i) - P(t_{i+\kappa j})\|^2
\label{eq:lstm_loss}
\end{equation}
where we define composition of $H_P$ with itself by,
\begin{equation}
\begin{aligned}
H_P^1 (P_i) &= H_P (P(t_i), \,P(t_{i-\kappa}), \,P(t_{i-2\kappa}), \,\hdots), \\
H_P^2 (P_i) &= H_P (H_P^1 (P_i), \, P(t_{i}), \,P(t_{i-\kappa}), \,\hdots), \\
H_P^3(P_i)&= H_P (H_P^2 (P_i), \, H_P^1 (P_i), \,P(t_{i}), \,\hdots)\,\,\hdots .
\end{aligned}
\label{eq:lstm_composition}
\end{equation}
The sum over index $i$ in Eq. \eqref{eq:lstm_loss} is taken over all initial times in the training dataset.  The same network structure and loss function is used for the full set of 50 pressure measurements as well as the 32 dimensional reduced order representations of the flow field.  Layer sizes are scaled to account for the difference in dimension between $P$ and $\psi / \xi$.

For any $\tau \geq 0$, we obtain the estimated forecast,
\begin{equation}
\hat{P}(t+\tau) = H_P^{m_\tau}(P_t),
\label{eq:LSTM_forecast}
\end{equation}
where $m_\tau = \tau / (\kappa dt)$.   Following the LSTM based forcast of $P/\hat{\psi}/\xi$ we may use $F_{P/\hat{\psi}/\xi}$ trained for zero lead time to evaluate $\hat{q}(t+\tau) = F_{P} \circ H_{P}^{m_\tau} (P_t)$ and likewise for $\hat{\psi} / \xi$.  Since prediction of $q$ with zero lead time is a much simpler problem, we use standard feed-forward neural networks in place of the LSTMs used for non-dynamic predictions with non-zero lead time.

\subsection{Error metrics}\label{subsec:error_metrics}

Each of the neural networks in this work is trained using the mean square error (MSE).  The MSE is differentiable and may be evaluated on subsets of the training data, allowing for the use of stochastic optimization schemes run on a graphics processing unit. 
However, the MSE may not be a good indicator of success in predicting extreme events.  
We have therefore adapted several extreme-event-tailored error metrics to compare the various methods considered in this work. 
Specifically, results are compared using the batch relative entropy loss, the maximum adjusted area under the precision-recall curve, and the extreme event rate dependent area under the precision recall curve, and optimal $F_1$ score.

The batch relative entropy loss (BRE) is inspired by the work in \cite{qi2020using}, where authors use a relative entropy loss function to train convolutional neural networks that are capable of making accurate prediction of a system governed by the truncated Korteweg-de Vries (tKdV) equation in regimes with extreme events.  
Their work uses empirical partition functions similar to the soft-max activation commonly used in neural networks to transform high-dimensional predictions into probability distribution functions highlighting outlier values.
Loss is subsequently measured loss using the KL-divergence.  
This approach was shown to significantly improve prediction accuracy over the MSE on the tKdV problem.
A similar approach was adapted for the present work using partition functions over mini-batches rather than output dimensions.  
Specifically, we define the batch relative entropy loss as,
\begin{equation}
BRE = \sum_i z_i \log\left( \dfrac{z_i}{\hat{z}_i} \right)
\end{equation}
where $z_i$, $\hat{z}_i$ are given by the empirical partition functions,
\begin{equation}
z_i = \dfrac{e^{q(t_i)}}{\sum_j e^{q(t_j)}}   ,  \hspace{1cm} \hat{z}_i = \dfrac{e^{\hat{q}(t_i)}}{\sum_j e^{\hat{q}(t_j)}},
\end{equation}
and where the sum is taken over a mini-batch.
We note that this work considers partition functions that weigh positive outliers, as opposed to the symmetric variant used in \cite{qi2020using}, since all events of interest to this work skew positive.  
The BRE loss was tested as a means of training the neural networks described in previous sections, but taken over mini-batches was found to perform comparably with the mean square error.
For evaluation, we use the batch relative entropy loss taken over the full testing dataset.

The performance of a predictor of extreme events may also be measured by the ability of that predictor to classify events based on a threshold value of the quantity of interest.
The maximum adjusted area under the precision-recall curve ($\alpha^*$) was proposed in \cite{guth2019machine} as a loss function for predicting extreme events and shown to perform well for predicting extreme dissipation events in Kolmogorov flow and rouge waves in the Majda-McLaughlin-Tabak model. 
To define $\alpha^*$ we first introduce the quantity $\omega$ for the extreme event rate and the corresponding threshold $\hat{a}(\omega)$, such that $p(q > \hat{a}(\omega)) = \omega$.  Introducing a second threshold $\hat{b}$ for the prediction $\hat{q}$ yields a classifier for which we can compute the precision ($S$ = true positives divided by predicted positives) and recall ($R$ = true positives divided by total positives).  Noting that precision is uniquely determined by the extreme event rate and recall, \cite{guth2019machine} computes the area under the precision recall curve,
\begin{equation}
\alpha (\omega) = \int_0^1 S (R,\omega) \, dR = \int_\bbR S(\hat{b},\omega)\left| \frac{\partial R(\hat{b},\omega)}{\partial \hat{b}} \right| \, d\hat{b},
\label{eq:AUC}
\end{equation}
and the maximum adjusted area under the precision-recall curve as 
\begin{equation}
\alpha^* = \underset{\omega}{max} \, \alpha (\omega)  - \omega.
\label{eq:alpha_star}
\end{equation}
We compute the integral in Eq.  \eqref{eq:AUC} using a finite grid of $\hat{\beta}$ values spread evenly between $min(\hat{q})$ and $max(\hat{q})$ with finite difference scheme to evaluate the derivative term.  It's value is bounded between $0$ and $1$ but occasionally falls slightly outside this range due to numerical issues and is subsequently clipped.  The maximization in Eq. \eqref{eq:alpha_star} is taken over a discrete set of samples evenly spaced $\omega$ between $0.01$ and $0.25$.

Finally, we consider the extreme event rate dependent optimal $F_1$ score, defined as the $F_1$ score on the testing dataset using the threshold that optimizes the $F_1$ score on the training and validation datasets. That is,
\begin{equation}
F_{1,opt}(q, \hat{q}, \omega) = F_1 \left( (q_{test} > \hat{a}(\omega)), (\hat{q}_{test} > \hat{b}_{opt}) \right) 
\label{eq:F1_opt}
\end{equation}
where,
\begin{equation}
\hat{b}_{opt} = \underset{\hat{b}}{arg max} \, F_1 \left( (q_{train/val} > \hat{a}(\omega)), (\hat{q}_{train/val} > \hat{b}) \right).
\end{equation}

Taken together this yields five metrics of predictor performance.  Mean square error, batch relative entropy, and $\alpha^*$ are independent of extreme event rate and yield simple scalar metrics of performance, though do not indicate performance at a particular extreme event rate.  The extreme event rate dependent $\alpha(\omega)$ and $F_{1,opt}(\omega)$ each seek to measure a balance between precision and recall at variable extreme event rates.  The dependency on extreme event rate allows for a more descriptive quantification of error, since we can compare methods for a variety of extreme event rates.

\section{Results} \label{sec:results}

In this section we present results for each of the methods for Eq. \eqref{eq:prediction_problem}. 
Results for prediction using the full order model with reduced order initial condition are kept distinct from those using neural networks. 
This is due to the weak performance by the former, as well as its considerable computational expense, which limits the number of trials we use to evaluate it.
For predictions of $q(t)$ using dynamic methods, we apply a smoother to remove higher frequency oscillations. 
This smoother weights previous predictions with exponentially desceasing weights and does not use any future prediction beyond $\tau$. 
Examples of full test set predictions for $\hat{q}(t)$ using each of the six nerual network based methods are shown in Figure \ref{fig:tau7_time_series} for lead time $\tau = 7.0$ and in Appendix D for other lead times.

\subsection{Simulations with reduced order initial conditions}

\begin{figure*}[h]
\centering
\includegraphics[width=\textwidth]{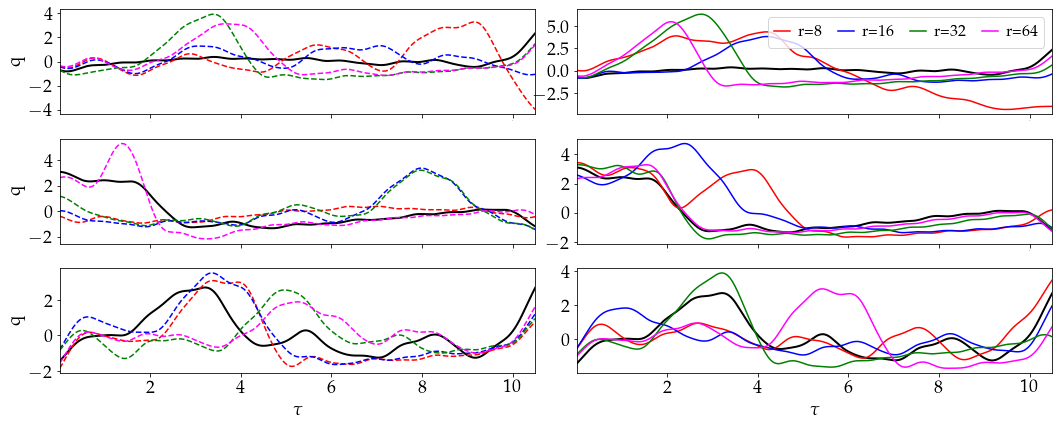}
\caption{Examples of forecasts of $q(t+\tau)$ using Eq. \eqref{eq:CFD_forecast} with various ranks of POD (left column) and FFNN (right column).  Examples show prediction from initial conditions are spaced evenly throughout the testing dataset.  True value is shown as solid black line. Each row represents a different realization of the flow. }
\label{fig:CFD_restart_examples}
\end{figure*}

We first present results showing the failure of approaches taken by the authors to forecast aerodynamic fluctuations using the Navier-Stokes solver and reduced order initial condition.
For each of a variety of ranks, reduced order initial conditions were formed using the estimated POD time series $\hat{\psi}(t)$ and FFNN flow reconstruction $D \circ E (P_t)$ at 50 evenly spaced times throughout the portion on data reserved for testing.  
Examples of the resulting smoothed drag coefficient are shown in Figure \ref{fig:CFD_restart_examples}, which compares predicted time series for $\hat{q}(t)$ using several ranks of POD reconstructions (left column) as well as FFNN based reconstructions (right column) to the true value $q(t)$ (black curves). 
Each row represents a different initial condition from within the testing dataset. 
We note that all the initializations using different reduction methods/orders exhibit poor agreement with the true time series in at least 1 of the three cases shown.

\begin{figure*}[h]
\centering
\includegraphics[width=\textwidth]{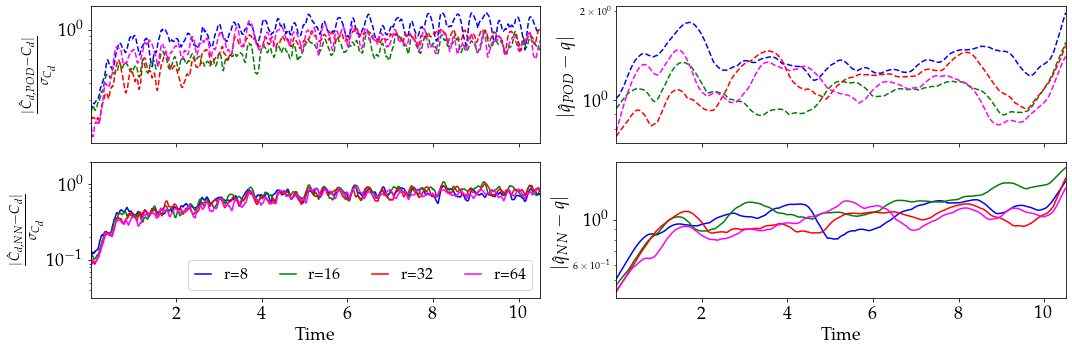}
\caption{Mean aboslute error, normalized by standard deviations of true time series for the drag coefficient and $q(t)$ using various ranks of reduced order initial conditions from the estimated POD time series and FFNN.}
\label{fig:CFD_restart_error}
\end{figure*}

The mean absolute error in the drag coefficient and smoothed drag coefficient $q$  for each rank are shown in Figure \ref{fig:CFD_restart_error}.
There are several noteworthy features;
the approximated values of $C_d$ for both methods are initially fairly accurate, falling within $10 \%$ of the true values, but very quickly diverge.
By the end of the $10.5$ time unit simulations, the error in each quantity is roughly as large as the standard deviation of the true data.
Initial error in $q(t)$ is higher, as is expected since this quantity includes some information from future estimates of $C_d$.
We also note the lack of noticable correlation between the rank of the latent space used for flow reconstruction and prediction accuracy.
In the POD case, it appears that the lowest rank ($r=8$) performs poorly compared to higher ranks, but $r=64$ is not uniformly better than $r=16$ or $r=32$.
A plausible explanation for this is that as the rank is increased, so too is the difficulty of the sensing problem outlined at the end of Sec. \ref{subsub:PODFR}.
Indeed, using the true (non-reconstructed) $\psi(t)$ we see improved agreement with full order results if $r$ is let to become large.
The connection between rank and accuracy for the FFNN examples is more opaque.  
While higher rank initial conditions are slightly more accurate at predicting the initial drag, the difference is small and decays quickly as the simulation progresses.

While the results here are not sufficiently exhaustive to preclude the use of a full order model for predicting extreme events in the flow around an airfoil they at least provide strong evidence of its difficulty.
We include them to show the sensitivity of the problem to small changes in the initial condition (note that the FFNN was able to accurately reconstruct the fluid velocity) and to contrast to other examples in fluids where a representation in low rank basis was found to be an effective precursor to extreme events \cite{blonigan2019extreme}.

\subsection{Comparison of reduced order dynamic methods}

\begin{figure*}[h]
\centering
\includegraphics[width=\textwidth]{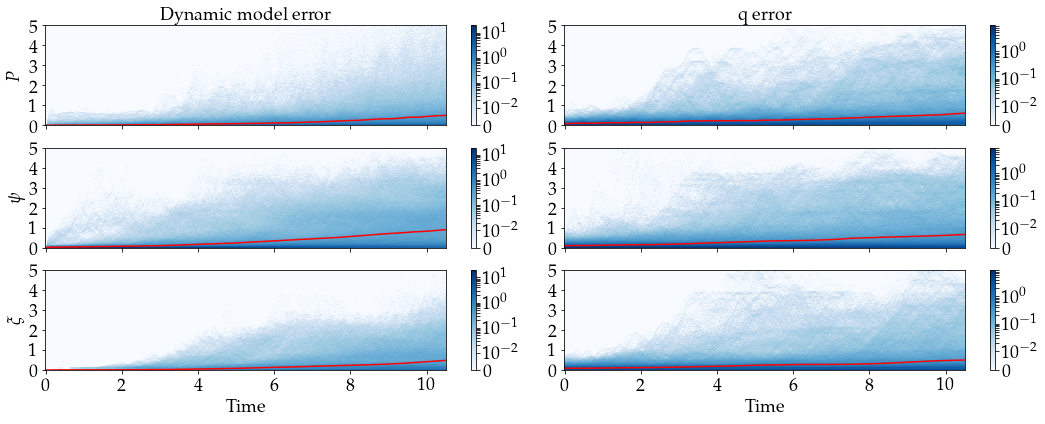}
\caption{Test set empirical probability density functions of the mean square error of the LSTM forecast (left) and absolute error in predicting $q(t+\tau)$ for $50$ dimensional data-driven forecast of pressure measurements (top), and $32$ dimensional reduced order models using POD modes (middle) and FFNN latent space (bottom).  Mean error quantities are shown by red curves.}
\label{fig:ROM_error_histograms}
\end{figure*}

We now consider the data-driven dynamic models for the pressure signals and reduced order representations of the flow field. 
Figure \ref{fig:ROM_error_histograms} shows the time evolution of the mean square error in the dynamic quantity and absolute error in prediction of $q$ for each of the three methods $H_{P/ \psi /\xi}$.  
The dynamic quantities in each case have been normalized to unit variance so direct comparison of error magnitudes across methods and across indices are meaningful. 
Error statistics have been binned for each timestep and normalized such that vertical slices of any subplot are density functions across all examples from the testing dataset.  Apparent recurring features in the error plots for $q$ are likely due to the methods missing the same feature from multiple closely sampled initial conditions.

In each of the three models, the bulk of the error remains low throughout the forecast window, as is shown by the curves for mean error.
Error statistics for the dynamic model for $P$ remain low for approximately three time units before some trajectories diverge.
Outlier errors for the dynamic model for $\psi$ grow more rapidly than others initially and by the end of the 10.5 convective unit interval are noticeably larger, on average than then other two.
The dynamic model for $\xi$ clearly has both the lowest error in the dynamic variable and has error for $q$ of a similar magnitude to the dynamic model in $P$.

\begin{figure*}[h]
\centering
\includegraphics[width=\textwidth]{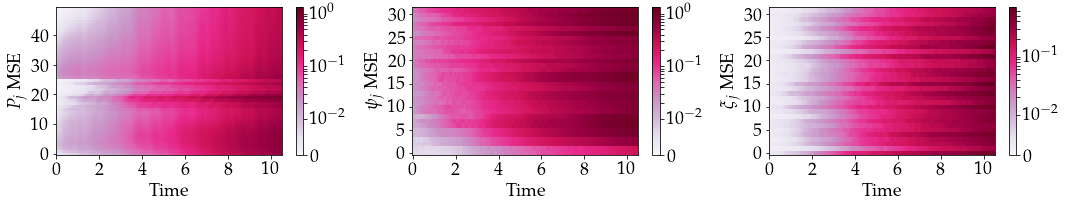}
\caption{Time evolution of mean (across test set samples) square error for each of 50 pressure sensors (left), 32 POD modes (middle) and dimensions of FFNN latent space (right) using LSTM based dynamic models.}
\label{fig:ROM_error_index}
\end{figure*}

For the models in $P$ and $\psi$ where indices in the dynamic variable carry meaning is also instructive to see what variables accumulate error at what rates. 
Figure \ref{fig:ROM_error_index} shows the mean square error for each index in each dynamic variable taken across test set examples.
The sharp discontinuity in the error for pressure sensors is a consequence of their position on the airfoil, shown in the lower panel of Fig. \ref{fig:grid_vort}. 
The first 26 pressure sensors are equally placed on the top of the airfoil from front to tail including endpoints and the subsequent 24 on evenly spaced from front to tail on the lower side.
We see that error towards the tail of the airfoil on both the suction and pressure sides is initially lower than towards the front. 
However, as time progresses, there is considerable error in pressure sensors towards the rear of the suction edge of the airfoil. 

As expected error in the low order POD modes is lower than in the high order, less energetic, modes. 
This may be explained by the fact that these low order modes tend to track large scale features of the flow and have slower and less chaotic trajectories. 
It is worth noting that, while the left hand column of Fig. \ref{fig:ROM_error_histograms} makes clear that error in $\psi$ grows much more rapidly in the initial forecast than $P$ or $\xi$, the same is not as obvious for low $\tau$ prediction of $q(t+\tau)$ using the dynamic model for $\hat{\psi}$.  
This suggests that the value of $q$ may largely be a function of the low order energetic POD modes, which Fig. \ref{fig:ROM_error_index} shows are accurately tracked for longer lead times.
Indices for the FFNN latent space variable $\xi$ do not have meaning and there is no correlation between index and the rate at which error increases.

\subsection{Forecasting aerodynamic fluctuations}

\begin{figure*}[h]
\centering
\includegraphics[width=\textwidth]{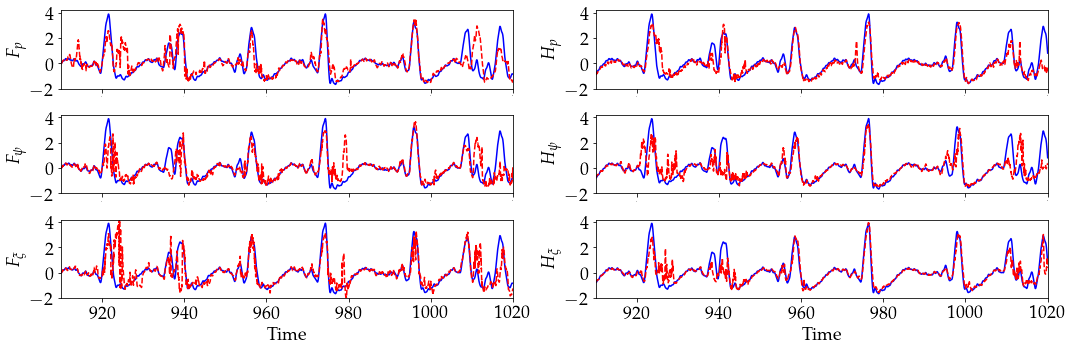}
\caption{True (blue) and predicted (red) time series for $q(t)$ with lead time $\tau=7$.}
\label{fig:tau7_time_series}
\end{figure*}

In this section we compare the six neural network based prediction strategies using the metrics outlined in Sec. \ref{subsec:error_metrics}. Examples of predictions for lead time $\tau=7.0$ using all methods are shown in Fig. \ref{fig:tau7_time_series}.  Blue curves indicate true values $q(t)$ and red dashed curves show predictions with lead time $\tau=7.0$ using the three non-dynamic and three dynamics methods.  Plots for other lead times are shown in Appendix D.

\begin{figure*}[h]
\centering
\includegraphics[width=\textwidth]{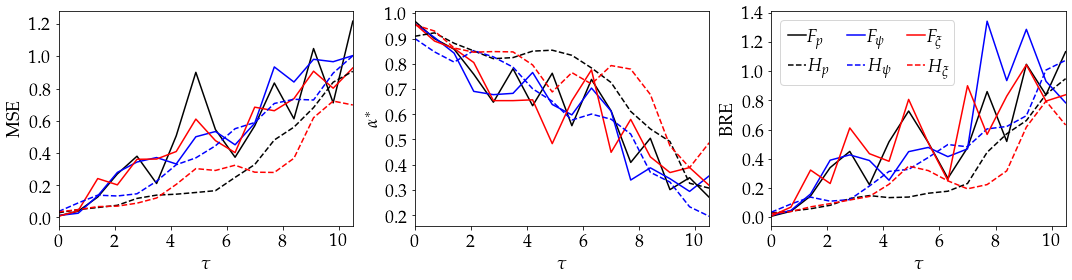}
\caption{Mean square error, maximum adjusted area under the precision-recall curve ($\alpha^*$), and test set relative entropy for each of the six neural network based prediction methods.  Evaluation is performed on partition of data reserved for method comparrison (test set).}
\label{fig:MSE_alpha_KL}
\end{figure*}

Scalar values (MSE, $\alpha^*$, BRE) of prediction accuracy for each method and lead times $\tau$ ranging from $0$ to $10.5$ convective time units in intervals of $0.7 \approx 1/f_{peak}$ are shown in Fig. \ref{fig:MSE_alpha_KL}. 
Metrics for prediction via non-dynamic methods are shown as solid lines and those for dynamic models are shown as dashed lines.

Figure \ref{fig:MSE_alpha_KL} clearly indicates some qualitative differences between dynamic and non-dynamic methods. 
In particular, error statistics for the dynamic models at a particular value of $\tau$ are highly correlated with those for similar lead times. 
The curves indicating their performance as a function of lead time are therefore smooth. 
The same is not true for non-dynamic models, where predictions for each lead time are performed via their own trained neural network. 
It is plausible that more care could be taken to yield consistency across lead times for non-dynamic models but methods for doing so are not immediately apparent. 

The scalar error metrics also clearly show that the dynamic models in $P$ and $\xi$, $H_{P}$ and $H_{\xi}$ are the most accurate of the six models.  Differences between the other four are less pronounce, though $H_\psi$ does, on average, slightly outperform non-dynamic methods.
This difference is most notable in the MSE, though still apparent for the extreme event specific metrics. 
This suggests that a non-negligible portion of the improvement $H_\psi$ has over non-dynamic methods manifests in the non-extreme values of $q$, though the scalar valued metrics do not resolve this feature.

\begin{figure*}[h]
\centering
\includegraphics[width=\textwidth]{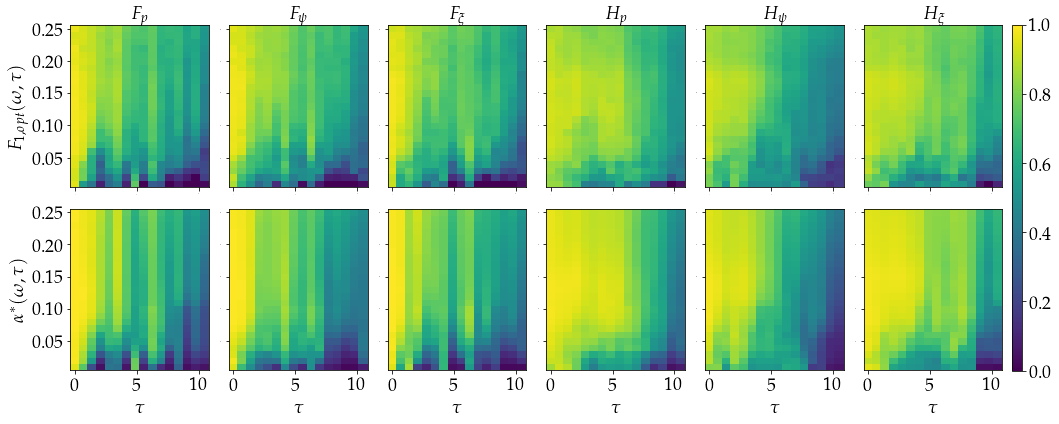}
\caption{Extreme event rate dependent area under the precision-recall curve ($\alpha(\omega)$ and $F_{1,opt}$) for each method evaluated on test set.}
\label{fig:F1_and_AUC}
\end{figure*}

Values for the remaining extreme event rate dependent evaluation metrics are shown in Fig. \ref{fig:F1_and_AUC} for values of $\omega$ between $0.01$ and $0.25$. 
The top row shows values for $F_{1,opt}$ and bottom row shows the area under the precision-recall curve. 
Within each subplot, values towards the bottom of the image indicate the prediction accuracy of the model for classifying very rare events $(1\%)$ while those at the very top show more common events, with lead time increasing across the horizontal axis.
While differences in values of prediction accuracy using Fig. \ref{fig:F1_and_AUC} may not be immediately clear, the same qualitative features seen in Fig. \ref{fig:MSE_alpha_KL} are again apparent.  In particular, errors for the dynamic methods are much smoother in time than those of the non-dynamic methods.

\begin{figure*}[h]
\centering
\includegraphics[width=\textwidth]{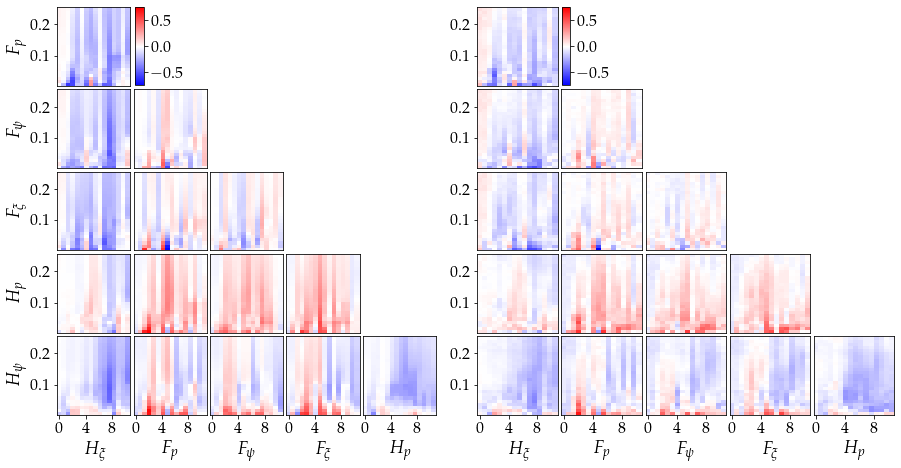}
\caption{Comparrison of $\alpha$ (left) and $F_{1,opt}$ (right) across various methods, lead times $\tau$, and extreme event rates $\omega$.  Each plot shows $\alpha/F_{1,opt}$ for method corresponding to row minus $\alpha/F_{1,opt}$ for method corresponding to column.   Blue (negative) indicates column-method outperforms row-method.  Hence lower left plot indicates $H_\xi$ has higher $\alpha$ (favorable performance) than $H_\psi$ in most cases. }
\label{fig:F1_AUC_comp}
\end{figure*}

Differences between methods become much more apparent when plotted directly.  Figure \ref{fig:F1_AUC_comp} shows the differences in values between all methods for both $F_{1,opt}$ (left) and area under the precision-recall curve, $\alpha$ (right).  Values within each subfigure indicate the metric evaluated with the method on the corresponding row minus the metric for the method on the corresponding column.  Hence, blue (negative) indicates the the method corresponding to that column performed better while red (positive) indicates the method corresponding to the row is better.

Figure \ref{fig:F1_AUC_comp} shows $H_P$ and $H_\xi$ outperforming other methods across all lead times $\tau$ and extreme event rates $\omega$ except for a small number of datapoints clustered around $\tau=4$ and low $\omega$. 
Comparisons between other methods are less easily summarized.
There does not appear to be discernible trends in the comparison between non-dynamic methods using $P$, $\psi$ or $\xi$ as input.
From this it seems reasonable to surmise that the accuracy of non-dynamic methods is not improved in a meaningful manner by exploiting flow field data.
The dynamic model for $\psi$ outperforms non-dynamic methods for mid-range lead times and lower extreme event rates, but under perform in prediction of less rare events (higher $\omega$) at longer lead times.
Differences between methods are largely consistent between $F_{1,opt}$ and $\alpha$.

\section{Discussion}\label{sec:discussion}

We have investigated the feasibility of a variety of methods for forecasting aerodynamic fluctuations occurring in a simulation of two dimensional incompressible flow around a NACA 4412 airfoil using only pressure recordings along the surface of the airfoil. 
Some of these methods also exploited, in an offline manner, knowledge of the flow field to construct low dimensional representations that were either used as inputs to machine learned predictors or variables in dynamic models.
Comparison between the methods considered in this work suggests that the complexity of the flow, though low Reynolds and only two dimensional, precludes the use of a full order computational fluid dynamic model with low dimensional input. 
Other methods were largely comparable, except for the learned dynamic models of the pressure measurements and of the latent variable of the FFNN based flow reconstruction, which performed notably better than others.

The authors highlight that neural networks allow for tremendous freedom with regards to structure,  hyper-parameters,  regularization,  and other factors that may affect performance.
We have tested several architectures and selected the best result for this paper,  but acknowledge that our work falls considerably short of testing across all sizes and training procedures.
Doing so would require an extensive computational resources and results would have no guarantee of generality beyond the problem specifically discussed in this work.
Rather, we sought to investigate the use of various classes of models (non-dynamic, ROM, full order simulation) and representations of data (POD modes, FFNN) for the purpose of forecasting a quantity exhibiting extreme events.
We also note that the work contained in this manuscript has been done without the inclusion of artificial noise, as is common in works applying machine learning to synthetic datasets.
The authors suspect that including noise would not change the results in a meaningful way, since flow reconstruction methods have been shown to be robust to noise \cite{maulik2020probabilistic,erichson2020shallow} and quantities with dynamic models may be estimated with filtering.
Nonetheless, further work studying the effect of noise would be necessary to confirm this.

In Section \ref{sec:results} we suggested that low order initializations of the Navier Stokes solver failed to accurately predict fluctuations, this motivated the use of data-driven dynamic models. 
This is not surprising, given the complexity and non-periodic nature of the flow.
It is possible that a mixed strategy of projecting the governing Navier-Stokes equations onto POD modes coupled with an LSTM closure model would outperform both approaches, as suggested in \cite{levine2021framework}. 
This approach, however would be highly non-trivial when using the latent space of the FFNN, since it is unclear how one might project the known governing equations.

We also stress the nature of this work as a computational study, rather than one that may be directly applied to engineering.
Numerical simulation for the current work was implemented with a blockage ratio of $3.18\%$.  
This falls within the range that might be expected for experiments on bluff bodies \cite{yarusevych2011vortex, west1982effects} but in lower than some works studying the aerodynamic properties of airfoils. 
Several works have noted dependence of aerodynamic properties including the drag coefficient, Strouhal frequency certain critical Reynolds numbers on the blockage ratio \cite{turki2003effect}. 
This effect was observed in the case used for the present work, but not explored in depth.
We also note that the simulation is of two dimensional flow over a smooth airfoil, which exhibits qualitative differences from three dimensional simulations and experiments \cite{tank2017possibility}.  
Thus, the present work should be considered as a study of a computational model of prototypical flow phenomena, rather than experimental or application conditions.

The present work considered the case where pressure measurements are taken at 50 positions around the perimeter of the airfoil.
This is, of course, not practical and future work could consider the use of various sensor placement techniques for determining optimal placement on the airfoil with respect to forecasting fluctuations \cite{le2020deep, de2017enkf}.

We believe this work establishes compelling comparisons and baselines for the prediction of extreme events using measurements on the surface of an airfoil. 
Numerous techniques, including several adapted from other works on extreme events were tested and compared.
The results suggest that the use of data-driven dynamic models for quantities subsequently used to predict extreme events outperform those that ignore dynamics.
This work also provides numerous opportunities for future research. 
In particular, sensor placement and uncertainty quantification are critical elements of practical engineering that may be explored in the context developed in this work.

\section*{Acknowledgments} 
This material is based upon work supported by the Army Research Office (Grant No. W911NF-17-1-0306) and used the Extreme Science and Engineering Discovery Environment (XSEDE) \cite{towns2014xsede}, which is supported by National Science Foundation undder Grant No. ACI-1548562. TPS has also been supported by the Air Force Office of Scientific Research (Grant No. MURI FA9550-21-1-0058).  Simulations were performed on Expanse at the San Diego Supercomputing Center through allocation TG-MTH210003.

\small
\begin{spacing}{.5}
\bibliographystyle{plain}
\bibliography{refs}
\end{spacing}

\normalsize

\newpage
\section*{Appendix A: Numerical simulation details}\label{app:simulation_details}

Numerical simulations of the airfoil used in this work were performed using the open-source spectral element solver Nek5000 \cite{nek5000-web-page} and run on Expanse at the San Diego Supercomputing Center using a grant through the National science Foundation funded Extreme Science and Engineering Discovery Environment (XSEDE) \cite{towns2014xsede}.
The spectral element method, proposed in \cite{patera1984spectral}, parititions the computational domain into non-overlapping elements and using polynomial interpolation within each element to represent the solution.
Mesh was generated using gmsh \cite{geuzaine2009gmsh}.

Time integration was performed using a second order semi-implicit scheme described in \cite{fischer2003implementation}.
The operator-integrating-factor splitting method proposed in \cite{maday1990operator} was used to allow for stable time integration with a fixed timestep of $0.001$.
The filtering method proposed in \cite{fischer2001filter} was also used to stabilize the simulation.

The simulation was initialized with a velocity of $\bu = (1,0)$.
This lead to a short transient which was avoided when training neural networks and taking the proper orthogonal decomposition.
The inflow boundary condition was held at a fixed value of $\bu = (1,0)$ throughout the simulation. 
Boundary conditions on the wall of the airfoil were no slip (Dirichlet), along the top and bottom on the domain were symmetric .
The outflow boundary condition was of the convective type proposed in \cite{dong2015convective}.
This allowed for the passage of strong vortices out of the domain while avoiding numerical issues.

Simulation for 1020 convective units took approximately 3.5 hours running on 128 cores or approximately 5 days running on 16 cores on the author's local computer. 
Aerodynamic quantities are saved every timestep and pressure at discrete points on the surface of the airfoil is saved every 10 steps, or 0.01 time units.
Velocity and pressure data on the full domain are saved every 0.25 time units.
Code for recreating datasets, as well as files for time series of pressure and aerodynamic coefficients are available online at \url{https://github.com/snagcliffs/Airfoil_EE}.

\section*{Appendix B: Neural network structures and implementation}\label{app:NN_details}
All neural networks used in this work were implemented in Python using the TensorFlow library \cite{tensorflow2015-whitepaper}.  
The swish activation function \cite{ramachandran2017searching} was used between fully connected non-recurrent layers and the sigmoid function was used for in LSTM layers.  
In section \ref{subsec:xi_rom} we note that prediction of $q$ following LSTM dynamic models are performed using non-recurrent networks.
This step, which does not forecast any quantity, is a fairly trivial problem and details of networks have been omitted here but are available in the online code repository.

\subsection*{Structure:}\label{app:NN_structure}
Structures for neural networks used in this work are given in Eq. \eqref{eq:NN_structure}.  We use the notation $(\text{input shape}) \to (\text{layer type , layer sizes})$ with multiple integer layer sizes indicating repeated layers. 
Hence, the first line tells us $F_p$ uses 140 history points of dimension 50 as input, then a 32 dimensional fully connected layer, two LSTM layers of size 32, and finally a series of fully connected layers having sizes 32, 16, 8, 4, and 1.
The final layer of size 1 is the putput layer.
For networks including a latent representation of the dynamics, the quantity $r$ has been used a stand in for rank. 
This work includes examples having $r \in \{8,16,32,64\}$.
The output size of $E$ is the number of interpolation points in the spectral element grid $n=279,552$, multiplied by the dimension $d=2$, for a total size of $n\cdot d = 559,104$.
In all cases using history terms, a stride of 3 was used between inputs.  This was found to yield a slight performance advantage over using a stride of 1.
Connections in the decoder network, excluding those to the final large layer we modeled after residual type networks, having a linear connection added to those feeding into nonlinear activation functions.
A larger number of history points were used for $F_p$ to account for the history terms used to compute $\hat{\psi}$ and $\xi$, though this was not found to significantly affect results.
\begin{equation}
\begin{aligned}
&F_p:\, (50 \times 140) \to (FC, \, 32) \to (LSTM, \, 32, 32) \to (FC, \, 32,16,8,4, 1)\\
&F_\psi:\, (r \times 70) \to (FC, \, 32) \to (LSTM, \, 32, 32) \to (FC, \, 32,16,8,4, 1)\\
&F_\xi:\, (r \times 70) \to (FC, \, 32) \to (LSTM, \, 32, 32) \to (FC, \, 32,16,8,4, 1)\\
&G:\, (50 \times 70) \to (FC, \, 64) \to (LSTM, \, 128) \to (FC, \, 64,r)\\
&E: \, (50 \times 70) \to (FC, \, 64) \to (LSTM, \, 64) \to (FC, \, 64,r)\\
&D: \, (r \times 1) \to (FC, \, 64, 128, 256, n\cdot d) \\
&H_{p}:\, (50 \times 70) \to (FC, \, 50) \to (LSTM, \, 100, 100) \to (FC, \, 100,50,50)\\
&H_{\psi/\xi}:\, (r \times 70) \to (FC, \, 32) \to (LSTM, \, 64,64) \to (FC, \, 64,32,32)\\
\end{aligned}
\label{eq:NN_structure}
\end{equation}
As noted in section \ref{sec:discussion}, designing neural networks allows tremendous freedom on seemingly arbitrary choices such as layer sizes and activation functions.  When testing different networks for this work we found significant differences between network architectures (LSTM, fully connected, branch-trunk, etc.).  Small changes such as slightly altering layer sizes or activation functions did not, in general, significantly affect results.

\subsection*{Data partitions:}
In many machine learning tasks and particularly those with a high dimensional parameter space such as neural networks, a dataset is split into distinct sets for training, validation,  and testing.
The training dataset is used to update model parameters according to the gradient of the cost function.
The validation set is used to prevent overfitting through the use of early stopping when performance metrics on validation data have stalled.
Finally, the testing set is reserved for a comparrison between models.
We use a (70/15/15\%) split, meaning 70 percent of the available data in each case in used for training, 15 for validation, and 15 for testing. 
Training and validation datasets are mixed randomly from within an interval of time spanning the initial 85\% of the training data and testing data is the remaining contiguous 15\%.
The length of data available for training models exhibits slight variability due to input sequence lengths and lead times. 
Comparison between models in section \ref{sec:results} is therefore performed on the final 15000 datapoints, or equivalently 150 convective units of the simulation.

\subsection*{Training procedure:}

Parameters for the training procedures used for each neural network are given in Table 1.
We used the Adam optimizer \cite{kingma2014adam} with batch size as given in the last column.
Restarts, set to 3 for all networks except $E/D$, indicates the number of optimizations from random initial weights that were performed.  Each was run until validation error failed to decrease for a specified number of epochs, called the patience. 
The encoder-decoder pair was only trained once due to computational expense.
The quantities $\ell^{1/2}$-reg describe the $\ell^1$ and $\ell^2$ regularization weights applied to connections in non-recurrent layers. 
LSTM layers as well as connections to the output layer in each network were left unregularized.
We note that considerably inflated values used for $F_\xi$ are due to significant overfitting observed at lower regularization.
A pair (0.98,2) for decay indicates that learning rate was multiplied by 0.98 every two epochs.
The initial learning rate in each case was set to $0.001$.

\begin{tabular}{|p{1.4cm}||p{1.4cm}|p{1.4cm}|p{1.4cm}|p{1.4cm}|p{1.4cm}|p{1.4cm}|  }
 \hline
 \multicolumn{7}{|l|}{Table 1: Training Parameters for Neural Networks } \\
 \hline
Network & Restarts & $\ell^1$ reg. &  $\ell^1$ reg. & Decay & Patience & Batch     \\
 \hline
$F_P$   & 3 & 0 &  1e-3 & 0.98, 2 & 10 & 1000 \\
$F_\psi$   & 3 & 0 &  1e-3 & 0.98, 2 & 3 & 1000 \\
$F_\xi$   & 3 & 1e-2 &  1e-2 & 0.98, 2 & 3 & 1000 \\
$H_P$   & 3 & 0 &  1e-3 & 0.95, 2 & 5 & 250 \\
$H_\psi$   & 3 & 0 &  1e-3 & 0.95, 2 & 5 & 250 \\
$H_\xi$   & 3 & 0 &  1e-3 & 0.95, 2 & 5 & 250 \\
$G$ & 3 & 0 & 1e-5 & none & 20 & 100 \\
$E/D$ & 1 & 0 & 1e-3 & 0.95,1 & 5 & 10 \\
 \hline
\end{tabular}\\

Parameters for training procedures were largely set based on empirical evidence and considerations for computational resources. 
While we claim to have put forward due diligence in tuning all networks for the sake of a valid comparison, we make no claim that these values represent the optimal set for the problem at hand. 
The authors are not aware of any convincing methods for optimizing such hyper-parameters.

\section*{Appendix C: Wake instability}\label{app:instability}

\begin{figure*}[h]
\centering
\includegraphics[width=\textwidth]{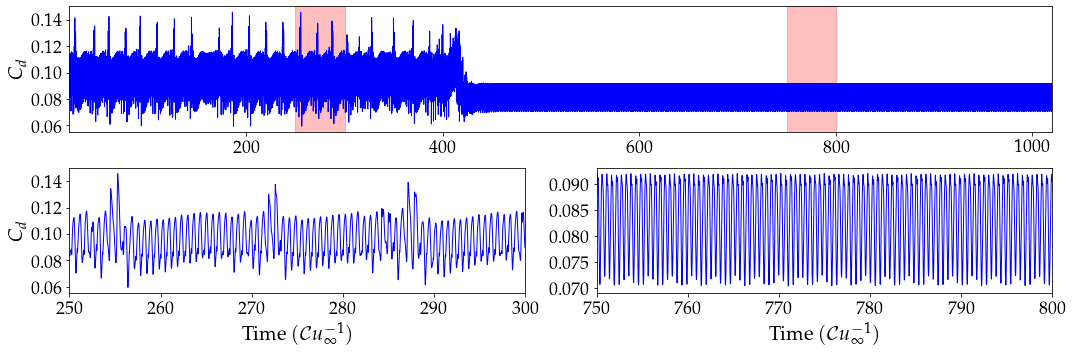}
\caption{Time series for the drag coefficient showing transition for intermittent to more regular behavior.  Lower row shows close up of time series from two shaded regions.}
\label{fig:unstable_drag}
\end{figure*}

\begin{figure*}[h]
\centering
\includegraphics[width=\textwidth]{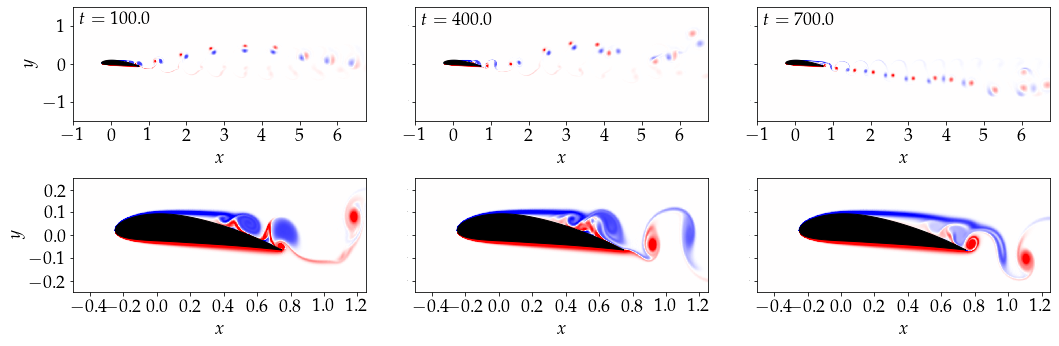}
\caption{Vorticity snapshots of the simulation exhibiting mode switching at time $t=100$ before switching, $t=400$ soon before switching, and $t=700$ after switching.}
\label{fig:unstable_vort}
\end{figure*}

In section \ref{sec:problem_description} we noted that the intermittent behavior examined in this work is quasi-stable and that in some cases a change in the wake pattern occurs, resulting in less chaotic behavior.  
To illustrate this, we include here a trajectory similar to the one considered in this work where this mode switching does occur.  
Figure \ref{fig:unstable_drag} shows $C_d(t)$ for a simulation exhibiting a shift in wake behavior around $t=400$.
Plots of the voritcity that clearly illustrate the transition are shown in \ref{fig:unstable_vort}.
Wake patterns before and after transition resemble the $P$ and $2S$ behaviors discussed in \cite{gopalakrishnan2017airfoil} and also shown in \cite{menon2020aerodynamic} at lower Reynolds numbers and higher angle of attack.

\section*{Appendix D: Time series of predicted quantity of interest using all methods}\label{app:all_time_series}

Time series prediction of $\hat{q}$ compared to the true value for test set examples at various values of lead time $\tau$ are shown in Fig. \ref{fig:all_time_series}. 
Networks used to generate these time series are described in further detail in Appendix B. 
The reduced rank representation of $\bu$ using both POD and FFNN use $r=32$.
Figures showing time series result for $\tau$ sampled between 0 and 10.5 in increments of 0.7 may be found online in the code repository for this work.

\begin{figure*}[h]
\centering
\includegraphics[width=\textwidth]{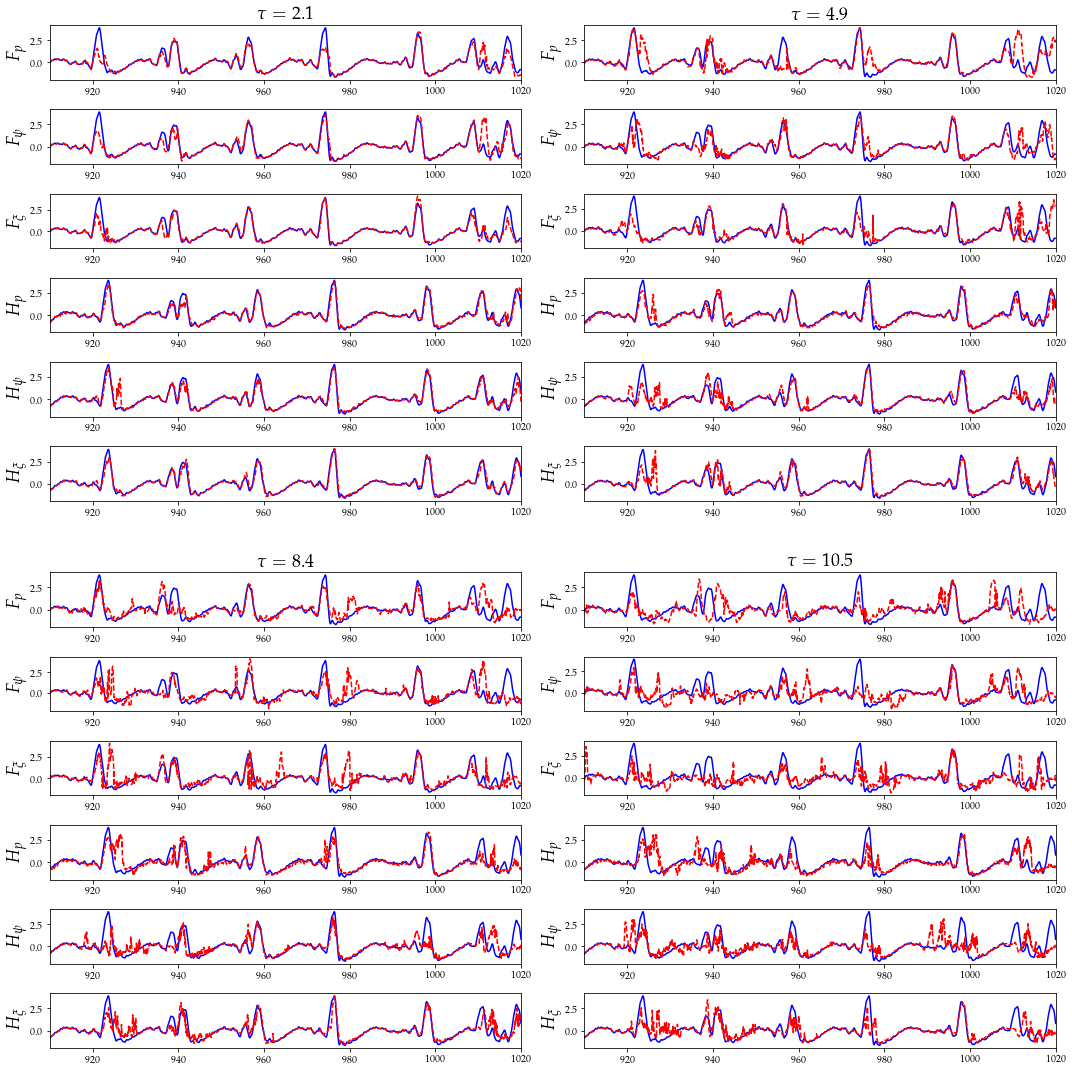}
\caption{Time series for $q(t)$ and $\hat{q(t)}$ using each of the neural network based methods on the test set for lead times $\tau = 1.4$, $3.5$, $7,0$, and $10.5$.}
\label{fig:all_time_series}
\end{figure*}

\end{document}